\documentclass[a4paper,11pt]{article}
\usepackage{pos}
\usepackage{multirow}

\begin{document}
\title{The splashback radius of groups and clusters of galaxies at low
redshifts}

\author*[a]{ F.G. Kopylova}
\author[a]{ A.I. Kopylov}

\affiliation[a]{SAO RAS, Nizhny Arkhyz, Russia}

\emailAdd{flera@sao.ru}
\emailAdd{akop@sao.ru}

\abstract{
We present a study of the distribution of galaxies along the radius
of 157 groups and clusters of galaxies
(200~km~s$^{-1}$ < $\sigma$ < 1100~km~s$^{-1}$)
of the local Universe  (0.01 < $z$ < 0.1).
We introduced a new boundary of galaxy systems
and identified it with the splashback radius $R_{sp}$.
We also identified the central region of galaxy systems with
a radius of $R_c$.
These radii are defined by the observed integrated distribution of the
total number of galaxies depending on the squared distance
from the center of the groups/clusters coinciding, as a rule, with
the brightest galaxy. We show that the radius $R_{sp}$ is proportional
to the $R_{200c}$ (radius of the virialized region of a galaxy cluster)
and to the radius of the central region $R_c$ with a slope close to 1.
Among the obtained dependences of the radii on X-ray luminosity, the
$\log R_{sp}$ - $\log L_X$ relation has the lowest scatter.
We measured $<R_{sp}>$ = $1.67\pm0.05$~Mpc for the total sample,
$<R_{sp}>$ = $1.14\pm0.14$~Mpc for galaxy groups with
$\sigma \leq$ 400~km~s$^{-1}$, $<R_{sp}>$ = $2.00\pm0.20$~Mpc for galaxy
clusters with $\sigma$ > 400~km~s$^{-1}$.
We found the average ratio of the radii $R_{sp}/R_{200c} = 1.40\pm0.02$ or
$R_{sp}/R_{200m} = 0.88\pm0.02$.}

\FullConference{%
The Multifaceted Universe: Theory and Observations -  2022 (MUTO2022)\\
23-27 May 2022\\
SAO RAS, Nizhny Arkhyz, Russia\\}


\maketitle

\section{Introduction}
Clusters of galaxies are the largest gravitationally bound objects in
the Universe. They are collapsing structures known as dark matter halos.
Clusters of galaxies are continuously increasing in mass both as a result
of the merging with individual galaxies and smaller groups of galaxies,
and as a result of continuous infall of dark matter from the environment.
Clusters of galaxies do not have clear boundaries, which are often
determined by density contrast relative to the critical or average
density of the universe. Their evolution is considered within the
framework of the spherical collapse model in the expanding universe
(e.g. \cite{1972ApJ...176....1G,1973ApJ...186..481G}).

Using N-body simulations of the motion of particles of a dark
matter halo (galaxies), it was found that a significant
part of them (up to 50\%) located outside the virialized regions of
galaxy clusters (up to 2$R_{vir}$ or 2$R_{200}$) have already been inside
\cite{2000ApJ...540..113B,2004A&A...414..445M,2005MNRAS.356.1327G}.

\cite{2015ApJ...806..101H} presents the results
of Millennium simulations for 75 galaxy clusters ($z$ = 0.0), where it
is shown that a significant part of the galaxies bounce up to
$3r_{proj}/r_{200}$ in the phase-space diagram.

The radius of galaxy clusters (physical halo boundary),
the splashback radius $R_{sp}$, was introduced in \cite{2014JCAP...11..019A}
as the radius at which newly accreted dark matter particles are piled
up within the apocenters of their orbits.
The $R_{sp}$ radius is clearly visible on the dark matter halo density
profiles as a sharp density drop \cite{2014JCAP...11..019A,2014ApJ...789....1D}.
In simulations performed in \cite{2015ApJ...810...36M}, it is shown that the
localization of $R_{sp}$ depends on the rate of mass accretion into the
cluster: in the halo with a rapid accretion rate
$R_{sp} = \sim 0.8$ - $1R_{200m}$,
in the halo with a slow rate --- $\sim 1.5R_{200m}$.

In this work, we are looking for observational manifestations of
splashback features in a sample of groups and clusters of galaxies
(data from the SDSS catalog). In \cite{2015AstBu..70..243K} we show the
edge of the
galaxy clusters, clearly identified by the integral distribution
of the number of all galaxies in a cluster depending on the squared
clustercentric distance, which we call the radius of the halo, $R_h$.
This radius is usually larger than the radius $R_{200c}$ and is measured
along the projected profile when a sharp increase in the number of
galaxies in the center of clusters ends. We identified them
later with the splashback radius $R_{sp}$ and gave the results of its
measurements for $\sim$~100 groups/clusters of galaxies
\cite{2015AstBu..70..243K,2016AstBu..71..257K,2018AstBu..73..267K,
2019AstBu..74..365K}.
In the works \cite{2018AstBu..73..267K,2019AstBu..74..365K} we have
shown that the
distribution  of early-type galaxies in clusters allows for a more
accurate estimate of the desired radius. We have measured for 40 galaxy
systems the average radius
$<R_{sp}> = 1.54\pm0.06~R_{200c}$ or $R_{sp} = 0.96\pm0.06~R_{200m}$
(if we take into account 4$R_{200c} \approx 2.5R_{200m}$),
which varies from 1.10 Mpc for the group NGC\,5627 with
$\sigma$ = 314~km~s$^{-1}$ to 4.17 Mpc for the cluster Coma (A\,1656)
with $\sigma$ = 921~km~s$^{-1}$. Here $R_{200c}$ (hereinafter $R_{200}$)
is the radius of a cluster inside which the density exceeds the critical
density of the universe by a factor 200. This radius in our works is
determined by the dispersion of radial velocities of galaxies in the systems.
In model simulations, the $R_{200m}$ radius is often used, within which
the density in the syatem exceeds the average density of the universe
by a factor 200.

In this study, we used a sample of 157 groups/clusters of galaxies
from the regions of superclusters of galaxies Leo (N=12), Hercules (N=27),
Ursa Major (N=19), Corona Borealis (N=8), Bootes (N=13), from other
smaller superclusters (N=11) and fields (N=20),
groups of galaxies from the region of the A\,1656/\,A1367
supercluster (N=48). For these systems of galaxies we have determined
radii $R_{sp}$ (splachback radius) and $R_c$ (core radius) from the
observed profile
and found the dependences of the radii on other galaxy cluster characteristics.
This study uses the data of the SDSS (Sloan Digital
Sky Survey Data Releases 7, 8) and 2MASS XSC (Two-Micron ALL-Sky Survey
Extended Source Catalog) catalogs and NED (NASA Extragalactic Database).
Throughout this study we adopted the following values of cosmological
parameters:
$\Omega_m = 0.3$, $\Omega_{\Lambda} = 0.7$, $H_0 = 70$~km~s$^{-1}$ ípc$^{-1}$.

\section{Method and data}

In the papers \cite{2015AstBu..70..243K,2016AstBu..71..257K,2017AstBu..72..363K,
2019AstBu..74..365K} we presented
the dynamical characteristics for a region with a radius of $R_{200}$
for almost the entire sample of groups and clusters studied in this work.
This radius can be estimated by the formula
$R_{200} = \sqrt {3} \sigma/(10H(z)$~Mpc \cite{1997ApJ...485L..13C}.
Then, if a cluster can be considered to be virialized within this radius,
its mass $M_{200}$ can be computed by the formula
$M_{200} = 3G^{-1}R_{200}\sigma^{2}$, where $\sigma$ is the dispersion of the
line-of-sight velocities of the galaxies located within the $R_{200}$ radius,
and G is the gravitational constant. Thus, the measured mass of the
cluster is  $M_{200} \propto \sigma^3$. In our work, we first estimated
the average line-of-sight velocity $cz$ of the cluster and
its dispersion $\sigma$, and then use the inferred dispersion
to determine the $R_{200}$ radius. We then determine the number of galaxies
within this radius and redetermine $cz$, $\sigma$, $R_{200}$, etc.
We move from the cluster and determine iteratively the dispersion
of line-of-sight velocities of the galaxies and other parameters of
the clusters within this radius. We consider galaxies with
velocities deviating by more than $2.7\sigma$ from the mean
velocity of the group (see, e.g., \cite{2004A&A...414..445M}) as field
objects.

To find the radius $R_{sp}$, it is important to highlight the
outskirts of galaxy systems. For this purpose, we present Figure 1 which
describes in detail the structure and kinematics of galaxy cluster
A\,1318 (as an example). The panels of Figure 1 show:
a) deviations of line-of-sight velocities of cluster members and
field galaxies from the average radial velocity of the system
plotted as a function of the squared clustercentric radius;
b) the integrated distribution of the number of galaxies as a function of
the squared clustercentric distance;
c) location of galaxies in the sky plane in equatorial coordinates;
d) the histogram of the line-of-sight velocities of all galaxies
within the $R_{200}$ radius. The solid line shows the
Gaussian corresponding to the dispersion of line-of-sight velocities of
galaxies. The vertical lines show the radii: $R_{sp}$ (dashed-and-dotted),
$R_{200}$ (short dashed), $R_c$ (long dashed).

\begin{figure}[!ht]
\centering
\includegraphics[width=14cm]{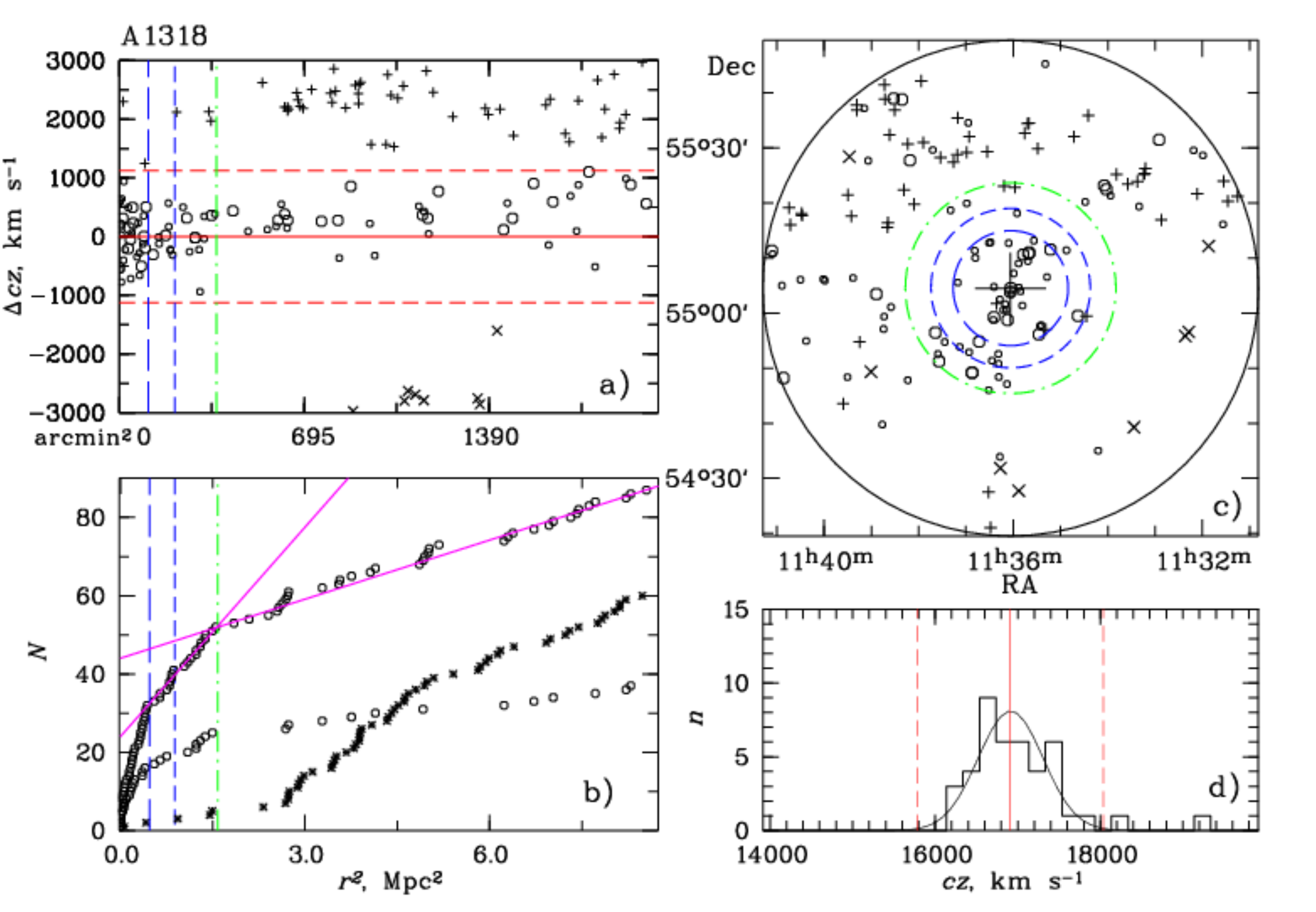}
\caption{
Distribution of galaxies in the cluster A\,1318.
The panel a) shows the deviation of radial velocities of galaxies from
the average cluster radial velocity determined from galaxies
within the radius $R_{200}$. The horizontal dashed red lines correspond
to $\pm2.7\sigma$ deviations, and the vertical blue short-dashed line marks
the $R_{200}$ radius; the blue long-dashed line marks the $R_c$ radius,
and the green dot-dashed line -- the $R_{sp}$ radius. The larger circles,
plus signs, and crosses mark the galaxies brighter than $M_K = -24^m$,
background galaxies, and foreground galaxies, respectively.
The panel b) presents the integrated
distribution of the total number of galaxies (the upper curve) as a function
of the squared projected distance from the cluster center. The lower
curve shows the distribution of early-type galaxies brighter than
$M_K<-21\,.\!\!^{\rm m}5$. The circles correspond to the galaxies denoted
by the circles in the top panel a), the crosses show the distribution
of field galaxies. The solid magenta lines show linear sections of the
profile. The panel c) shows the sky distribution (in quatorial
coordinates) of galaxies presented in the top panel a) (same designations
are used). The colored circles highlight the regions with radii $R_c$,
$R_{200}$, and $R_{sp}$. The studied region is bounded by a circle
of radius 3.5$R_{200}$ (solid black line). The large cross indicates the
center of the cluster.
The panel d) presents the distribution of radial velocities
of all galaxies within $R_{200}$ (the solid line shows the Gaussian for
cluster members). The solid vertical line indicates the
average radial velocity of the cluster, and the dashed lines correspond to
$\pm2.7\sigma$ deviations.
}
\label{clusA1318}
\end{figure}

We are especially interested in panel b), where the projected cluster
profile is shown, the integrated distribution of the number of galaxies
as a function of the squared clustercentric distance.
This distribution makes it possible to visually reveal the dense
core of the cluster, its more tenuous shell, and the external
region where the distribution becomes linear (shown by the straight
magenta lines in the figure) in the adopted coordinates,
i.e., where the distrubution of surrounding galaxies becomes uniform
on the average \cite{2015AstBu..70..243K}. The figure shows
the radius of the virialized region, $R_{200}$, and
the radius, $R_{sp}$, beyond which the steep growth of the cluster
members ends and linear growth begins. We also marked with a long
dashed line
the central part (core) of the cluster of radius $R_c$, where
the main steep increase in the number of galaxies is observed.
The lower curve in the same figure shows the distribution of early-type
galaxies brighter tnan $M_K=-21\,.\!\!^{\rm m}5$, which we use to refine
the radius in question. Such galaxies are located, as a rule,
in the central virialized regions of galaxy systems.
$R_{sp}$, the splashback radius of the galaxy systems
\cite{2014JCAP...11..019A,2014ApJ...789....1D}
found by us, is the radius of the apocenters of the orbits of galaxies,
originating in the central region of galaxy clusters.
Thus, the radius $R_{sp}$ separates galaxies which fall onto the cluster
for the first time from collapsing galaxies that are already participating in
establishing virial equilibrium.
For our sample  we measured $<R_{sp}> = 1.67\pm0.05$ Mpc
with a variation range of $0.75\div4.24$,
$<R_c> = 0.78\pm0.03$~($0.30\div2.00$) and
$<R_{sp}/R_{200c}> = 1.40\pm0.02$, or
$<R_{sp}/R_{200m}> = 0.88\pm0.02$ (given that $4R_{200c} \approx 2.5R_{200m}$.
This value is consistent with results of simulations (see, e.g.,
\cite{2015ApJ...810...36M}).

\section{Results}

\begin{figure}[ht]
\centering
\includegraphics[width=14cm]{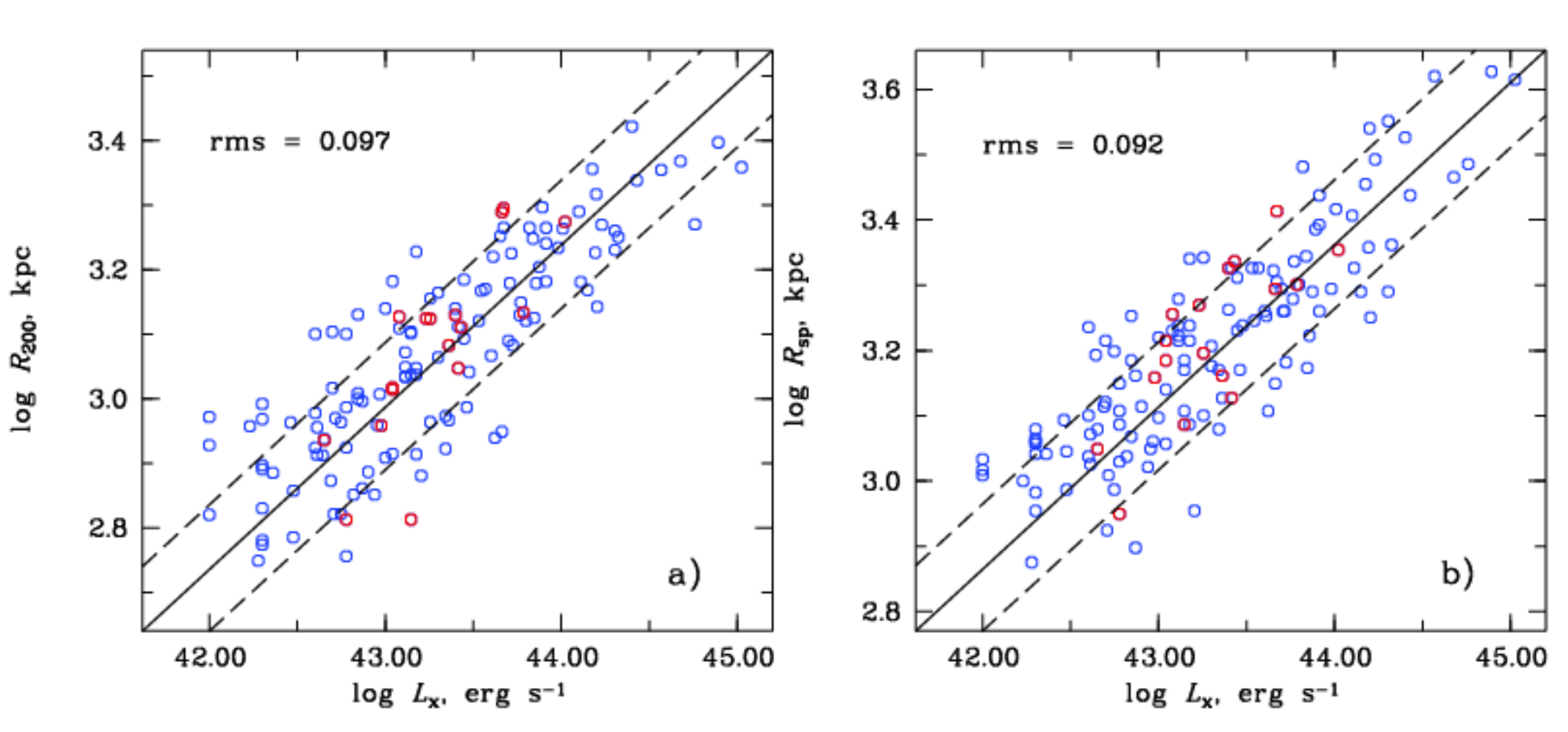}
\caption{The radii $R_{200}$ (a) and $R_{sp}$ (b) as functions of the X-ray
luminosity. The solid lines correspond to the relations
$R_{200} \propto L_X^{0.25}$ and $R_{sp} \propto L_X^{0.24}$.
Dashed lines correspond to $1\sigma$ deviations from them.
Red circles show groups/clusters of galaxies with bimodal
distribution of radial velocities.
}
\label{RLX}
\end{figure}

We stutied the dependences of the found radii $R_{sp}$ and $R_c$ on the
properties of galaxy clusters. Figure \ref{RLX} shows the dependence
of $\log R_{sp}$ on the X-ray luminosity,  $\log L_X$, and for comparison
a similar dependence for the radius $\log R_{200}$ is shown.
The relations shown (straight lines) represent the average between
forward and inverse regressions, when the independent variables are
interchanged. The dashed lines show 1$\sigma$ deviations from them.
Note that rms deviation depends on the $\log R_{sp}$ radius
less than on $\log R_{200}$. Red circles show merging clusters of
galaxies with bimodal radial velocity distribution within the radius
$R_{200}$. We can be see that these syatems do not differ from normal
clusters of galaxies by location on the $\log R_{sp}$--$\log L_X$ relation.
The Table~\ref{tab:data1} shows the parameters of the relations we obtained:
slopes, normalizations, and scatters.

We have noted in Section 1 the results of other authors
(e.g., \cite{2015ApJ...810...36M}), from which it follows that
$R_{sp}$ depends on the rate of accretion of dark matter particles on the
clusters: at a rapid rate of mass accretion, this radius is close to
the virial radius, that is, in our case to $R_{200}$.

\begin{figure}[ht]
\centering
\includegraphics[width=14cm]{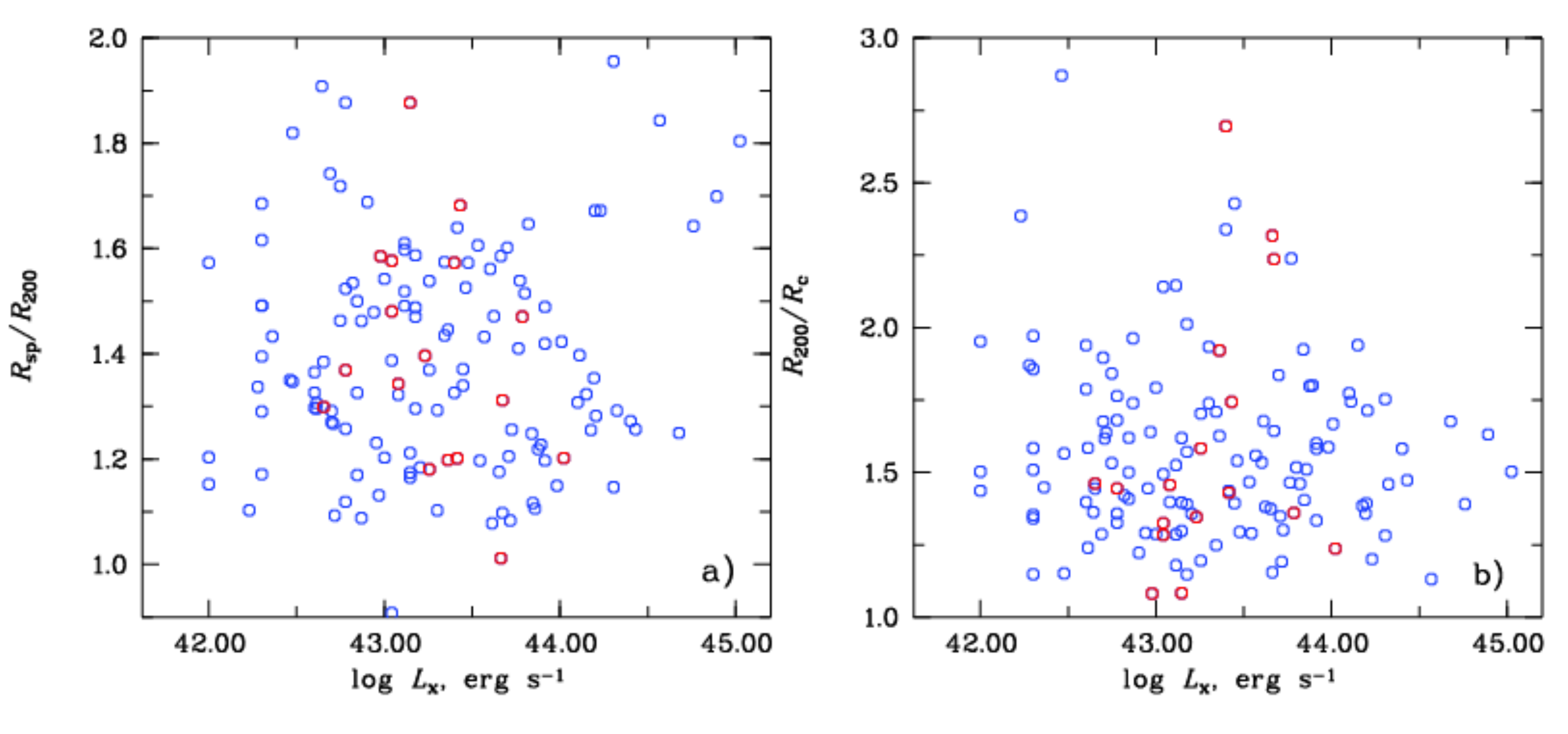}
\caption{The ratios $R_{sp}/R_{200}$ (a) and $R_{200}/R_{c}$ (b)
as functions of X-ray luminosity. Red circles show groups/clusters
of galaxies with bimodal distribution of radial velocities.
}
\label{RSR2}
\end{figure}

Figure~\ref{RSR2} shows the dependence of the radii ratios
$R_{sp}/R_{200}$ and $R_{200}/R_c$ on the X-ray luminosity of systems
of galaxies. Variations of the $R_{sp}/R_{200}$ ratio have distinct
boundaries for most systems of galaxies:
$R_{sp}/R_{200}$ = $1.2 \div 1.6$ and
$\log L_X$ = $42.5 \div 44.5$.
We refer to galaxy systems with a radius ratio less than 1.15
(or $R_{sp}/R_{200m} < 0.71$) as objects with a rapid rate of mass
accretion (RA), and the systems with a radii ratio
greater than 1.6 (or $R_{sp}/R_{200m} > 0.99$) - as objects with a slow
accretion rate (SA). There are  21 RA and 34 SA systems in our sample.
RA systems are usually groups/clusters of galaxies with a non-Gaussian
radial velocity distribution, with signs of merging with other groups
and galaxies near the virial radius, for example:
A\,1270, A\,1904, A\,1991, NGC\,2563.
Among SA groups/clusters of galaxies there are rich galaxy systems
such as A\,1656, A\,1795, A\,2142, A\,2029,
poor galaxy systems such as NGC\,7237, IC\,2476, MCG-01-29,
which collect matter (groups, galaxies, gas) from great distances
from the center.

We have obtained the following results:

1. The boundary of the dark halo of groups/clusters of galaxies, the radius
$R_{sp}$ determined by galaxies, is proportional to the radius of
the virialized region $R_{200}$ and to the radius of the core region $R_c$
with a slope close to 1.

2. All the radii we measured correlate with the X-ray luminosity
of groups/clusters of galaxies and have similar slopes.
Dependencies of the splashback radius on mass $M_{200}$ and
luminosity $L_{K,200}$ have a lower scatter.

\begin{table}[!th]
\caption{\label{tab:data1}Best fit parameters}
\centering
\begin{tabular}{|l|c|r|r|l}
\hline
Relation              & Slope  &  Normalization & Scatter \\
\hline
$\log R_{sp}$ -- $\log L_X$               & $0.24\pm0.03$ & $-7.39\pm0.33$ & 0.092 \\
$\log R_{200}$ -- $\log L_X$              & $0.25\pm0.04$ & $-7.60\pm0.34$ & 0.097 \\
$\log R_c$ -- $\log L_X$                  & $0.26\pm0.04$ & $-8.45\pm0.37$ & 0.110 \\
$\log R_{sp}$ -- $\log M_{200}/M_{\odot}$ & $0.32\pm0.02$ & $-1.42\pm0.13$ & 0.066 \\
$\log R_c$ -- $\log M_{200}/M_{\odot}$    & $0.35\pm0.03$ & $-2.08\pm0.17$ & 0.086 \\
$\log R_{sp}$ -- $\log L_K/ L_{\odot}$    & $0.42\pm0.03$ & $-2.00\pm0.17$ & 0.074 \\
$\log R_c$ -- $\log L_K/L_{\odot}$        & $0.44\pm0.03$ & $-2.66\pm0.19$ & 0.088 \\
$\log R_{sp}$ -- $\log R_{200}$           & $1.00\pm0.04$ & $+0.17\pm0.11$ & 0.064 \\
$\log R_{sp}$ -- $\log R_c$               & $0.95\pm0.05$ & $+0.46\pm0.14$ & 0.088 \\
\hline
\end{tabular}
\end{table}

\begin{acknowledgments}
This research has made use of the NASA/IPAC Extragalactic Database
(NED, \url{http://nedwww.ipac.caltech.edu}),
which is operated by the Jet Propulsion Laboratory, California Institute of
Technology, under contract with the National Aeronautics and Space
Administration, Sloan Digital Sky Survey (SDSS, \url{http://www.sdss.org}),
which is supported by Alfred P. Sloan Foundation, the participant institutes
of the SDSS collaboration, National Science Foundation, and the United
States Department of Energy and Two Micron All Sky Survey (2MASS,
\url{http://www.ipac.caltech.edu/2mass/releases/allsky/}).
\end{acknowledgments}

\bibliographystyle{JHEP}
\bibliography{Kopylova22.bib}

\begin{thebibliography}{15}
\providecommand\natexlab[1]{#1}
\providecommand\JournalTitle[1]{#1}

\bibitem[{Adhikari} {et~al.}(2014)]{2014JCAP...11..019A}
{Adhikari}, S., {Dalal}, N., {Chamberlain}, R. T. 2014, JCAP, 11, 19

\bibitem[{Balogh} {et~al.}(2000)]{2000ApJ...540..113B}
{Balogh}, M.~L., {Navarro}, J.~F., {Morris}, S.~L. 2000, \apj, 540, 113

\bibitem[{Carlberg} {et al.}(1997)]{1997ApJ...485L..13C}
{Carlberg}, R.~G., {Yee}, H.~K.~C., {Ellingson}, E., {et al.} 1997, \apj, 485, L13

\bibitem[{Diemer} \& {Kravtsov}(2014)]{2014ApJ...789....1D}
{Diemer}, B. \& {Kravtsov}, A.~V. 2014, \apj, 789, 1

\bibitem[{Gill} {et al.}(2005)]{2005MNRAS.356.1327G}
{Gill}, S.~P.~D., {Knebe}, A., {Gibson}, B.~K. 2005, \mnras, 356, 1327

\bibitem[{Gott}(1973)]{1973ApJ...186..481G}
{Gott}, J.~R,III 1973, \apj, 186, 481

\bibitem[{Gunn} \& {Gott}(1972)]{1972ApJ...176....1G}
{Gunn}, J.~E., \& {Gott}, J.~R,III 1972, \apj, 176, 1

\bibitem[{Haines} {et al.}(2015)]{2015ApJ...806..101H}
{Haines}, C.~P., {Pereira}, M.~J., {Smith}, G.~P., {et al.} 2015, \apj, 806, 101

\bibitem[{Kopylov} \& {Kopylova}(2015)]{2015AstBu..70..243K}
{Kopylov}, A.~I., {Kopylova}, F.~G. 2015, Asrophysical Bulletin, 70, 243

\bibitem[{Kopylova} \& {Kopylov}(2016)]{2016AstBu..71..257K}
{Kopylova}, F.~G., {Kopylov}, A.~I. 2016, Asrophysical Bulletin, 71, 129

\bibitem[{Kopylova} \& {Kopylov}(2017)]{2017AstBu..72..363K}
{Kopylova}, F.~G., {Kopylov}, A.~I. 2017, Asrophysical Bulletin, 72, 363

\bibitem[{Kopylova} \& {Kopylov}(2018)]{2018AstBu..73..267K}
{Kopylova}, F.~G., {Kopylov}, A.~I. 2018, Asrophysical Bulletin, 73, 267

\bibitem[{Kopylova} \& {Kopylov}(2019)]{2019AstBu..74..365K}
{Kopylova}, F.~G., {Kopylov}, A.~I. 2019, Asrophysical Bulletin, 74, 365

\bibitem[{Mamon} {et al.}(2004)]{2004A&A...414..445M}
{Mamon}, G.~A., {Sanchis}, T., {Salvador-Sole}, E., {Solanes}, M.~J. 2004, A\&A, 414, 445

\bibitem[{More} {et al.}(2015)]{2015ApJ...810...36M}
{More}, S., {Diemer}, B., {Kravtsov}, A.~V. 2015, \apj, 810, 36
\end{thebibliography}

\end{document}